\documentclass[12pt]{article}
\usepackage{authblk}
\usepackage[bookmarksnumbered, colorlinks, plainpages]{hyperref}
\usepackage{amsmath, amsthm, amscd, amsfonts, amssymb, graphicx, color, booktabs}
\numberwithin{equation}{section}
\definecolor{email}{rgb}{0.00,0.00,0.84}
\begin{document}
\setcounter{page}{1}


\title{
\vspace*{-13mm}
\begin{flushright}
\normalsize
KEK-CP-401
\end{flushright}
\vspace*{3mm}
\large \bf 12th Workshop on the CKM Unitarity Triangle\\ Santiago de Compostela, 18-22 September 2023 \\ \vspace{0.3cm}
\LARGE Status and progress of lattice QCD}

\author[1,2,3]{Takashi Kaneko \thanks{takashi.kaneko@kek.jp}}
\affil[1]{
  \small
  High Energy Accelerator Research Organization (KEK), Ibaraki 305-0801, Japan
}
\affil[2]{
  SOKENDAI (The Graduate University for Advanced Studies),
  Ibaraki 305-0801, Japan
}
\affil[3]{
  Kobayashi-Maskawa Institute for the Origin of Particles and the Universe (KMI),
  Nagoya University, Aichi 464-8602, Japan
}

\maketitle

\begin{abstract}
We review recent progress from lattice QCD for the determination of the Cabibbo-Kobayashi-Maskawa matrix elements.
\end{abstract} \maketitle


\section{Introduction}

\noindent Lattice QCD provides first-principles and non-perturbative
estimate of hadronic matrix elements needed to determine
Cabibbo-Kobayashi-Maskawa (CKM) matrix elements.
Independent realistic simulations have been available for the so-called
``gold-plated'' processes, where the initial and final states involve
at most one hadron stable in QCD without bottom quarks~\cite{FLAG5}.

Simulations of bottom quarks need fine and large lattices,
and hence are very time-consuming,
because the lattice spacing $a$ and spatial lattice size $L$ must satisfy
a condition $a \! \ll \! m_b^{-1} \! \ll \! M_\pi^{-1} \! \ll \! L$
in order to control discretization and finite volume effects (FVEs).
Currently available simulations, therefore, employ unphysically small masses or
effective-theory-based actions to be matched with QCD for bottom quarks.
Another and more essential problem for non-gold-plated processes is that
on-shell matrix elements of interest are not straightforward to be extracted 
from correlation functions on finite-volume Euclidean lattices.

In this article,
we introduce selected examples of recent progress
with precise realistic simulations for gold-plated processes
and that with newly developed methods for non-gold-plated processes.
We refer to Refs.~\cite{Rev:AV:Lat22,Rev:TK:Lat22,Rev:Meinel:Lat23,Rev:Tsang+DellaMorte:EPJ}
for more comprehensive reviews.


\section{Cabibbo angle anomaly}

\begin{figure} [htb!]
\centering
 \includegraphics[width=0.45\textwidth]{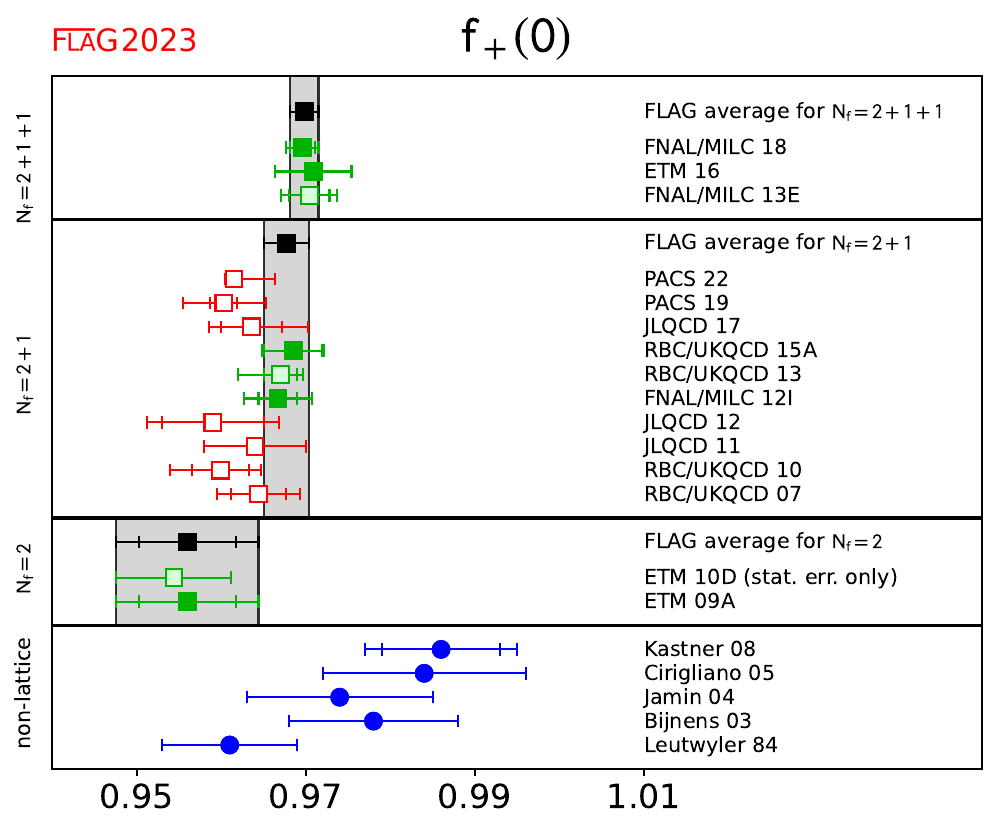}
 \hspace{0mm}
 \includegraphics[width=0.52\textwidth]{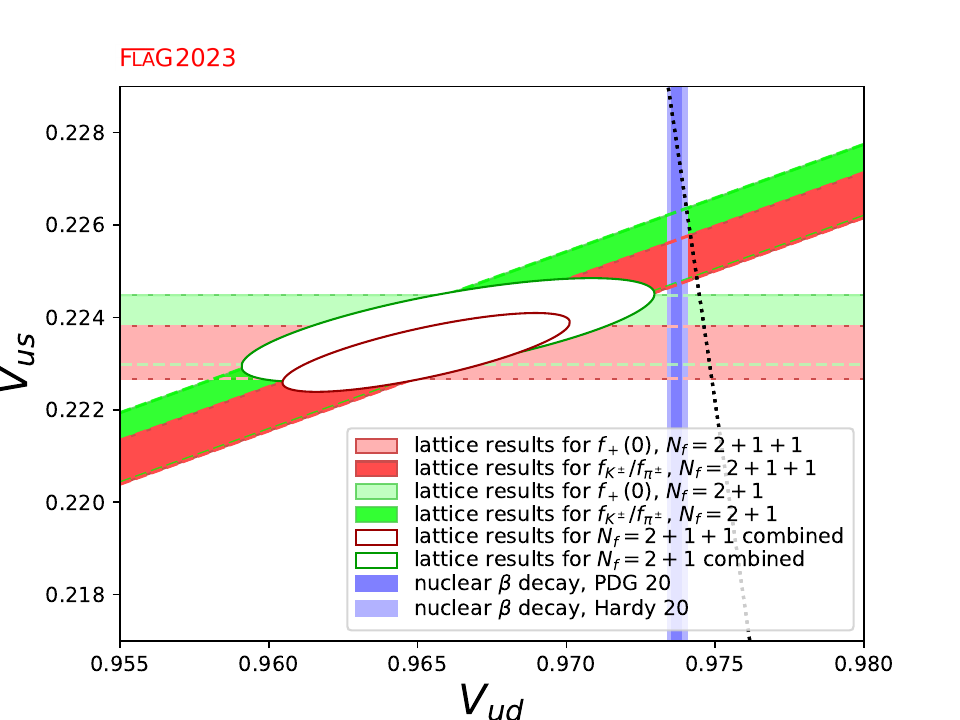}
 \vspace{-3mm}
 \caption{
   Left:
   lattice estimates of $K\to\pi$ form factor $f_+(0)$.
   These shown in green (red) squares (do not) satisfy FLAG's criteria
   to control systematics. 
   The black square and band for each $N_f$ show the average of the green squares.
   Blue symbols are phenomenological estimates.
   Right:
   $|V_{us}|$ versus $|V_{ud}|$.
   The slanted and horizontal bands show $|V_{us}|/|V_{ud}|$
   from kaon/pion leptonic decays and $|V_{us}|$ from $K\to\pi\ell\nu$ semileptonic
   decays, respectively.
   Red and green bands are estimates in $N_f=2+1+1$ and $2+1$ QCD, respectively.
   The blue bands show $|V_{ud}|$ from the nuclear $\beta$ decays.
   The black dotted line satisfies CKM unitarity.
   (Both panels from Ref.~\cite{FLAG5}.)
 }
\label{fig:kaon}
\end{figure}

Kaon leptonic and semileptonic decays are gold-plated,
and the condition for $a$ and $L$ is relaxed as 
$a \! \ll \! m_s^{-1} \! < \! M_\pi^{-1} \! \ll \! L$.
As shown in the left panel of Fig.~\ref{fig:kaon}, 
there have been independent and realistic simulations in this decade
leading to the sub-\% accuracy for hadronic matrix elements
to determine $|V_{us}|/|V_{ud}|$ and $|V_{us}|$,
namely, ratio of decay constants $f_K/f_\pi$
and $K\to\pi$ form factor at zero momentum transfer $f_+(0)$.
Together with a recent dispersive estimate of
universal radiative correction $1+\Delta_R$
to the superallowed nuclear $\beta$ decays~\cite{Delta_R:PRL,Delta_R:PRD},
a tension in CKM unitarity in the first row called ``Cabibbo angle anomaly''
has been reported, while it is alleviated by recent re-analysis of
nuclear dependent corrections~\cite{NuclDepCorr}.

The right panel of Fig.~\ref{fig:kaon} shows
recent estimate of $|V_{us}|/|V_{ud}|$ and $|V_{us}|$
by Flavour Lattice Averaging Group (FLAG)~\cite{FLAG5}
including ETM collaboration's result for $f_K/f_\pi$~\cite{fKfpi:ETM:Nf4}
after the last CKM workshop.
A measure of unitarity violation 
$\Delta_{\rm CKM}=|V_{ud}|^2+|V_{us}|^2+|V_{ub}|^2 - 1 = -0.0011(7)$
is consistent with zero,
when it is estimated from the kaon and pion leptonic and nuclear $\beta$ decays.
On the other hand, $\gtrsim 3\,\sigma$ tension is observed
with $|V_{us}|$ from the kaon semileptonic decays:
namely $\Delta_{\rm CKM}=-0.0021(4)$ together with nuclear $\beta$ decay
and $-0.0184(64)$ with the leptonic decays.
Since the average of $f_+(0)$ for $N_f=4$ is largely constrained
by a single precise study by Fermilab/MILC~\cite{f+0:FermilabMILC:Nf4},
independent calculations are welcome to firmly establish the lattice prediction
of $f_+(0)$.
Note also that, for $N_f=3$, 
the PACS collaboration reported their preliminary result in Ref.~\cite{f+0:PACS:Nf3}.

At the sub-\% accuracy of the hadronic inputs,
the uncertainty of the conventional calculation of radiative corrections
in chiral perturbation theory (ChPT) is no longer negligible.
The RM123-Soton collaboration developed a method to calculate
the radiative correction by decomposing the photon-inclusive decay rate
into its perturbative estimate for the point-like meson
and structure dependent correction on the lattice~\cite{iso:RM123Soton:method}.
Their latest result including the strong isospin correction 
$\delta_{\rm EM+SU(2)}=-1.26(14)$\,\%~\cite{iso:RM123Soton:Nf4}
is consitent with the conventional ChPT estimate
$-1.12(21)$\,\%~\cite{iso:ChPT:Cirigliano+}.
A recent indpendent calculation using a different lattice formulation also
obtained a consistent result $-0.86(^{+41}_{-40})$\,\%~\cite{iso:RBCUKQCD:Nf3}.
We also note that a different approach
using the so-called infinite volume recunstruction (IVR) method has been proposed,
and its applicability to the semileptonic decay is discussed
in Ref.~\cite{iso:Christ+:IVR}.
The IVR technique has been also applied to the first lattice calculation of
the $\gamma W$ box diagram responsible for the universal correction $1+\Delta_R$
to the nuclear $\beta$ decay~\cite{gammaW:l+pQCD}.
Their etimate $\Box_{\gamma W}^{VA, \leq 2\,\mbox{\scriptsize GeV}^2} \times 10^3= 1.49(8)$
is in good agreement with that from the dispersive analysis 1.62(10)
mentioned at the beginning of this section,
and is systematically improvable on finer lattices.


\section{$B$ meson exclusive semileptonic decays}

%
The conventional decay modes to determine $|V_{cb}|$ and $|V_{ub}|$,
namely $B \to D^{(*)}\ell\nu$ and $B \to \pi\ell\nu$, are gold-plated.
There has been steady progress in the lattice calculation of relevant form factors
to resolve the long standing tension in these CKM elements
with estimates from inclusive decays.
\begin{figure} [hbt!]
\centering
 \includegraphics[width=0.44\textwidth]{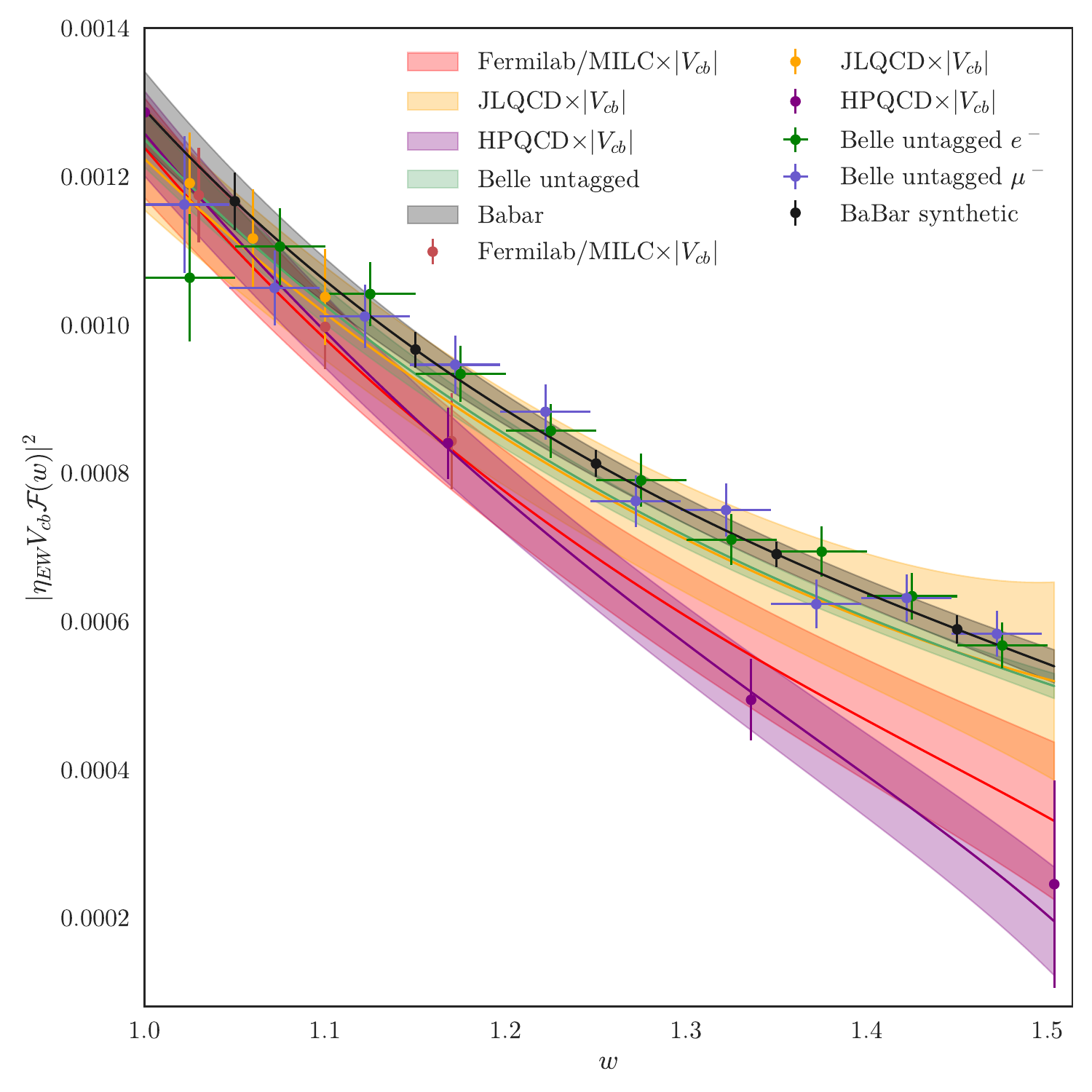}
 \hspace{15mm}
 \includegraphics[width=0.29\textwidth]{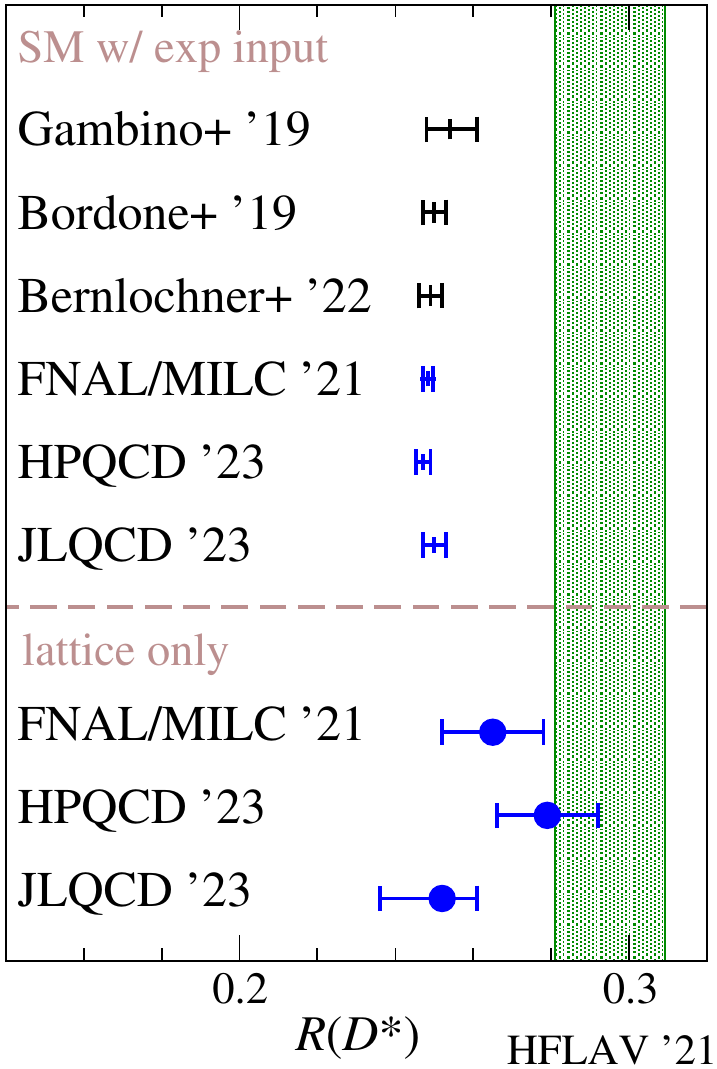}
 \vspace{-3mm}
 \caption{
   Left:
   comparison of recent lattice calculations of $B\!\to\!D^*\ell\nu$ form factors
   with experiments (figure from Ref.~\cite{BtoDstar:Vaquero:CKM23}).
   The differential decay rate without some kinematical factors,
   $|\eta_{EW} V_{cb} {\mathcal F}(w)|^2$,
   is plotted as a function of the recoil parameter $w$.
   The red, cyan and orange bands show results from
   Fermilab/MILC~\cite{BtoDstarFF:FermilabMILC},
   HPQCD~\cite{BtoDstarFF:HPQCD} and JLQCD~\cite{BtoDstarFF:JLQCD}, respectively,
   whereas 
   experimental data from Belle (BaBar) are plotted by the green (black) band.
   Right:
   recent estimates of $R(D^*)$. Pluses and circles are from analyses with
   and without experimental data, respectively.
   The vertical green band shows an average of experimental measurements~\cite{HFLAV21}.
 }
\label{fig:b2c}
\end{figure}
One of the most interesting progress is that
three collaborations have calculated  all $B \to D^*$ form factors,
$h_{A_1}$, $h_{A_2}$, $h_{A_3}$, $h_V$~\cite{BtoDstarFF:FermilabMILC,BtoDstarFF:HPQCD,BtoDstarFF:JLQCD,Colquhoun:CKM23},
while, until recently, only $h_{A_1}$ had been calculated only
at zero recoil~\cite{BtoDstarFF:FermilabMILC:w1,BtoDstarFF:HPQCD:w1}.
However, there is a tension among the new lattice calculations~\cite{Rev:AV:Lat22,BtoDstar:Vaquero:CKM23}
as shown in the left panel of Fig.~\ref{fig:b2c}, which shows
differential decay rate without some kinematical factors
$|\eta_{EW} V_{cb} {\mathcal F}(w)|^2$,
where $\eta_{EW}$ is the electroweak correction,
${\mathcal F(w)}$ is a functional of form factors given as
\begin{equation}
   {\mathcal F}^2
   \propto
   \left[
      2\frac{1-2wr+r^2}{(1-r)^2}
      \left\{ 1 + \frac{w-1}{w+1} R_1^2 \right\}
     +\left\{ 1 + \frac{w-1}{1-r} \left( 1 - R_2 \right) \right\}^2
   \right]
   h_{A_1}^2
\end{equation}
with a mass ratio $r=M_{D^*}/M_B$,
and form factor ratios
$R_1 = h_V/h_{A_1}$ and $R_2 = \left( rh_{A_2} + h_{A_3} \right)/h_{A_1}$.
The recoil parameter $w = v_Bv_{D^*}$ is defined 
using $B$ and $D^*$ meson four velocities.
Fermilab/MILC employed the so-called Fermilab action~\cite{Fermilab}
based on heavy quark effective theory (HQET) to directly simulate
physical bottom quark mass.
While their data (red band) shows slightly steeper slope than
experimental data (green and black bands),
these data can be fitted to a model independent parameterization
of $w$ dependence~\cite{BGL} with acceptable value of $\chi/{\rm dof} \sim 1.5$.
HPQCD employed a relativistic approach,
namely a relativistic lattice action (highly improved staggered quark action)
at unphysically small bottom quark masses to avoid large discretization erorrs,
and reported an even steeper slope (cyan band).
This also leads to a tension in the londitudinal polarized fraction for
the light lepton channels ($\ell = e, \mu$),
which is not easy to explain
even if new physics is introduced~\cite{b2c:NP,b2c:NP:CKM23}\footnote{
HPQCD's paper~\cite{BtoDstarFF:HPQCD} has been recently updated,
and show better consistency with other lattice and experimental data.}.
On the other hand,
JLQCD's data (orange band) obtained along a relativistic approach
with a chiral fermion action show good consistency in the kinemarical distribution
with experiment.
The source of this tension is not yet clarified.
Extenting their simulations safely to larger recoils
is a key to resolve the situation and for more extensive comparison with experiment.

The right panel of Fig.~\ref{fig:b2c} shows recent estimates of $R(D^*)$.
The previous phenomenological analyses of experimental data
already achieved 1\,\% accuracy with minimal lattice input $h_{A_1}(w=1)$,
and observed $\gtrsim 3\,\sigma$ tension with experiment~\cite{RDstar:Ph:Gambino+,RDstar:Ph:Bordone+, RDstar:Ph:Bernlochner+}.
A slightly better accuracy of $\gtrsim 0.5$\,\% is attained
using the recent lattice data of other form factors as well~\cite{BtoDstarFF:FermilabMILC,BtoDstarFF:HPQCD,BtoDstarFF:JLQCD}.
The same figure also shows purely theoretical estimates
without experimental input of the differential decay rate.
While the unsertainty increases to $\sim 5$\,\%
and the tension with experiment becomes unclear,
the accuracy is systematically improvable by more realistic simulations
particularly at large recoils.

\begin{figure} [hbt!]
\centering
 \includegraphics[width=0.31\textwidth]{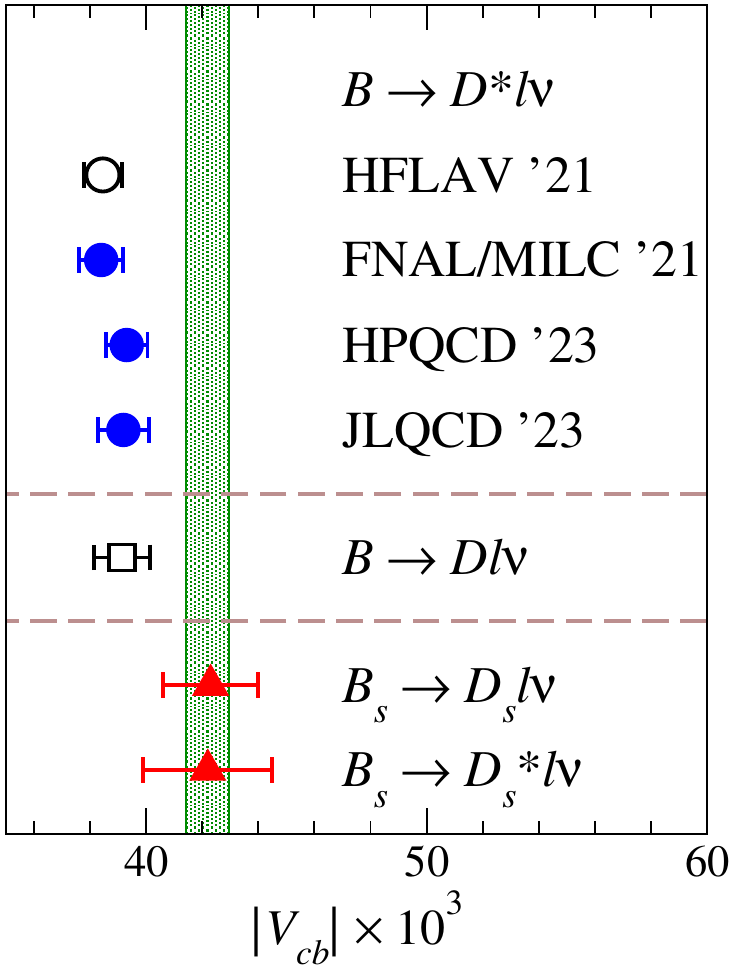}
 \hspace{15mm}
 \includegraphics[width=0.30\textwidth]{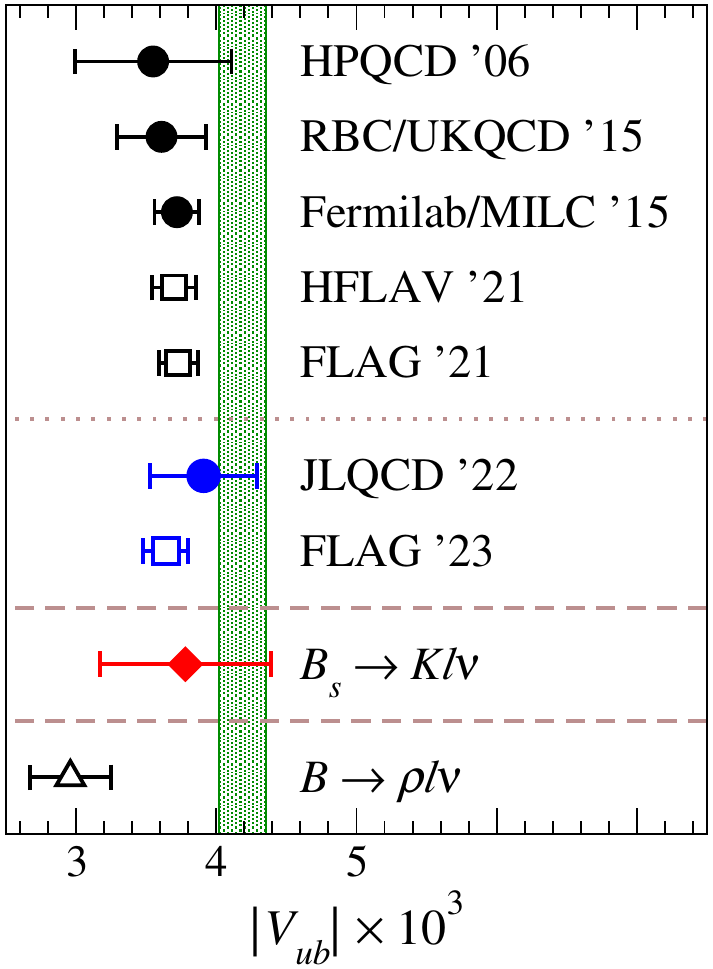}
 \vspace{-3mm}
 \caption{
   Left:
   recent estimate of $|V_{cb}|$.
   Circles, square and triangles are obtained from the $B \to D^* \ell\nu$,
   $B \to D \ell\nu$ and $B_s$ meson decays, respectively.
   The open circle shows previous estimate using a minimal lattice input $h_{A_1}(1)$,
   whereas blue circles are from recent lattice calculations of all form factors.
   Right: 
   recent estimate of $|V_{ub}|$.
   Black and blue circles show previous and JLQCD's results, respectively,
   whereas black and blue squares are world averages using these data.
   We plot estimates with recent lattice calculation of
   the $B_s \to K\ell\nu$ form factors (red diamond)
   and phenomenological estimate of $B \to \rho\ell\nu$ form factors (triangle).
   In both panel, the vertical bands show results from the inclusive decays.
 }
\label{fig:Vcb+Vub}
\end{figure}
The left panel of Fig.~\ref{fig:Vcb+Vub} shows recent estimate of $|V_{cb}|$.
The conventional determinations from $B \to D^*\ell\nu$ (open circle)
and $B \to D\ell\nu$ (open square) are consistent with each other,
and sigificantly smaller than that from the inclusive decay (green band).
HPQCD also calculated the form factors for $B_s \to D_s^{(*)}\ell\nu$,
although resulting $|V_{cb}|$'s are consistent both with
those from the exclusive and inclusive $B$ decays
within their 5\,\%uncertainty~\cite{BstoDs:HPQCD,BstoDsstar:HPQCD}.
The recent three lattice calculations of all form factors obtained
$|V_{cb}|$ in good agreement with the previous determination
in spite of the tension of their form factor data.
We note that the HPQCD value is obtained from a simultaneous analysis of
$B \to D^*\ell\nu$ and $B_s \to D_s^*\ell\nu$ decays,
and they obtain a rather large value of $45.5(2.0) \times 10^{-3}$,
if they use $\Gamma(B \to D^*\ell\nu$) alone.
The $|V_{cb }|$ tension is, therefore, remains unsolved.
It has been suggested that the $|V_{cb}|$ determination may suffer from
a bias called D'Agostini effecs~\cite{RDstar:Ph:Gambino+,DAgostini}.
While a model independent parametrization of $w$ dependence is available~\cite{BGL},
how to safely apply the Bayesian prior to poorly determined fit parameters is another
subtle issue~\cite{Bayesian:Flynn+,Bayesian:Flynn+:Lat23}.

Recent estimates of $|V_{ub}|$ are shown in the right panel of Fig.~\ref{fig:Vcb+Vub}.
JLQCD recently calculated the $B \to \pi\ell\nu$ form factors,
where the largest uncertainties come from the statistics
and the ``chiral extrapolation'' from their simulated pion masses $M_\pi \gtrsim 230$~MeV
to the physical mass~\cite{Btopi:Nf3:JLQCD,Colquhoun:CKM23}.
This is why rather old Fermilab/MILC study
with smaller $M_\pi \gtrsim 165$\,MeV
achieved a better accuracy.
World averages of $|V_{ub}|$ has been largely constrained
by this study for about 10 years,
and hence 
new independent calculations of the $B \to \pi$ form factors are higly welcome.

It is known that the $B_s \to K\ell\nu$ decay is advantageous in the control of
the statistical accuracy~\cite{Lepage} and chiral extrapolation to the physical $M_\pi$.
Figure~\ref{fig:Vcb+Vub} also shows $|V_{ub}|$
from a recent lattice study of the relevant form factors~\cite{BstoK:RBCUKQCD}
combined with LHCb data of branching fractions
${\mathcal B}=(B_s \to K\ell\nu)/(B_s \to D_s\ell\nu)$~\cite{BstoK:LHCb}
and $\Gamma(B_s \to D_s\ell\nu)$~\cite{BstoDs:LHCb}.
It is, however, consistent both with estimates
from the exclusive and inclusive $B$ decays within its 16\,\% uncertainty.
More precise experimental and lattice results are necessary
to provide an alternative determination of $|V_{ub}|$
competitive to the conventional one from $B \to \pi\ell\nu$.

We also plot $|V_{ub}|$ obtained fom $B \to \rho\ell\nu$
using a light-cone sum rule estimate of relevant form factors~\cite{Btorho:LCSR}.
It is not ``gold-plated'' with unstable $\rho$ in the final state,
and hence the lattice study is not straightforward.
Preliminary results based on a newly developed finite volume framework
are presented at this workshop~\cite{Btorho:Leskovec:CKM23}.
It is interesting to extend this study to the $B \to K^*\ell\ell$ decay,
for which a tension between the Standard Model and experiment
has been reported~\cite{P5prime}.


\section{Inclusive decays}

Lattice QCD can provide first principles estimate of $B$ meson matrix elements
needed in the conventional calculation of the inclusive decay rate
based on the operator product expansion (OPE).
However, a direct lattice calculation of the inclusive rate is also
attractive subject, 
since it enables a systematic comparison between the exclusive and inclusive
determinations of $|V_{ub}|$ as well as $|V_{cb}|$
in the same simulation to resolve their tensions.
The hadronic tensor $W_{\mu\nu}$ 
describing non-perturbative QCD effects to, for instance, $B \to X_c\ell\nu$
apears in the spectral decomposition of $B$ meson four-point function
on the lattice as
\begin{equation}
  C_{\mu\nu}(t,{\bf q})
  =
  \sum_{\bf x}
  \frac{e^{i\bf qx}}{2M_B}
  \langle
    \bar{B}({\bf 0}) \left| J_\mu^\dagger({\bf x},t) J_\nu(0) \right| B({\bf 0}) 
  \rangle
  =
  \int_0^\infty d\omega W_{\mu\nu}(\omega,{\bf q}) e^{-\omega t},
\end{equation}
where $J_\mu$ represents the weak current,
$(t,{\bf x})$ and $q$ are coordinates and momentum transfer,
respectively, and $\omega = M_B - q^0$ is the haronic final state energy.
It is, however, an ill-posed inverse problem to determine $W_{\mu\nu}$ from $C_{\mu\nu}$,
because $W_{\mu\nu}$ is largely distorted in a finite volume
due to the momentum direcretization:
namely, its multi-particle continuum part turns into
superposition of the $\delta$ function like singularities.
Recent breakthrough in direct lattice studies of inclusive processes is based on
the idea that, on the lattice, a smeared hadornic tensor\cite{smearedSF}
and its energy integral~\cite{energyInt:H} can be calculated to a good preision.
The latter developed into a direct calculation of the inclusive rate as 
\begin{equation}
  \Gamma
  =
  \frac{G_F^2 |V_{cb}|^2}{24\pi^3} 
  \int_0^{q_{\rm max}^2} d{\bf q}^2 \sqrt{{\bf q}^2} \bar{X}({\bf q}^2),
  \hspace{3mm}
  \bar{X}({\bf q}^2)
  = 
  \int_{\omega_{\rm min}}^{\omega_{\rm max}} d\omega K_{\mu\nu}(\omega,{\bf q}) W_{\mu\nu}(\omega,{\bf q})
  \label{eqn:incl}
\end{equation}  
with integration kernel $K_{\mu\nu}$ specified by the leptonic tensor
and kinematical factors for $\Gamma$, 
and by re-writing $\bar{X}$ (but not $W_{\mu\nu}$ itself) using $C_{\mu\nu}$~\cite{energyInt:GH}.
Here, ${\bf q}_{\rm max}^2$ ($\omega_{\left\{\rm min,max \right\}}$) represents
kinematically allowed region of ${\bf q}^2$ ($\omega$).

\begin{figure} [hbt!]
\centering
 \includegraphics[width=0.60\textwidth]{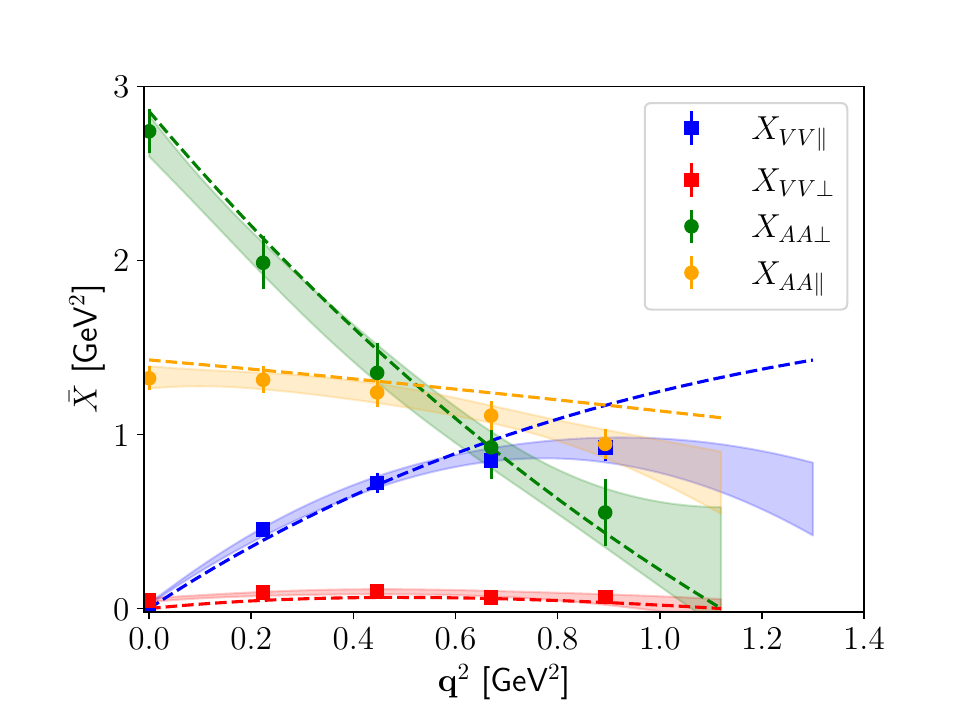}
 \hspace{5mm}
 \includegraphics[width=0.31\textwidth]{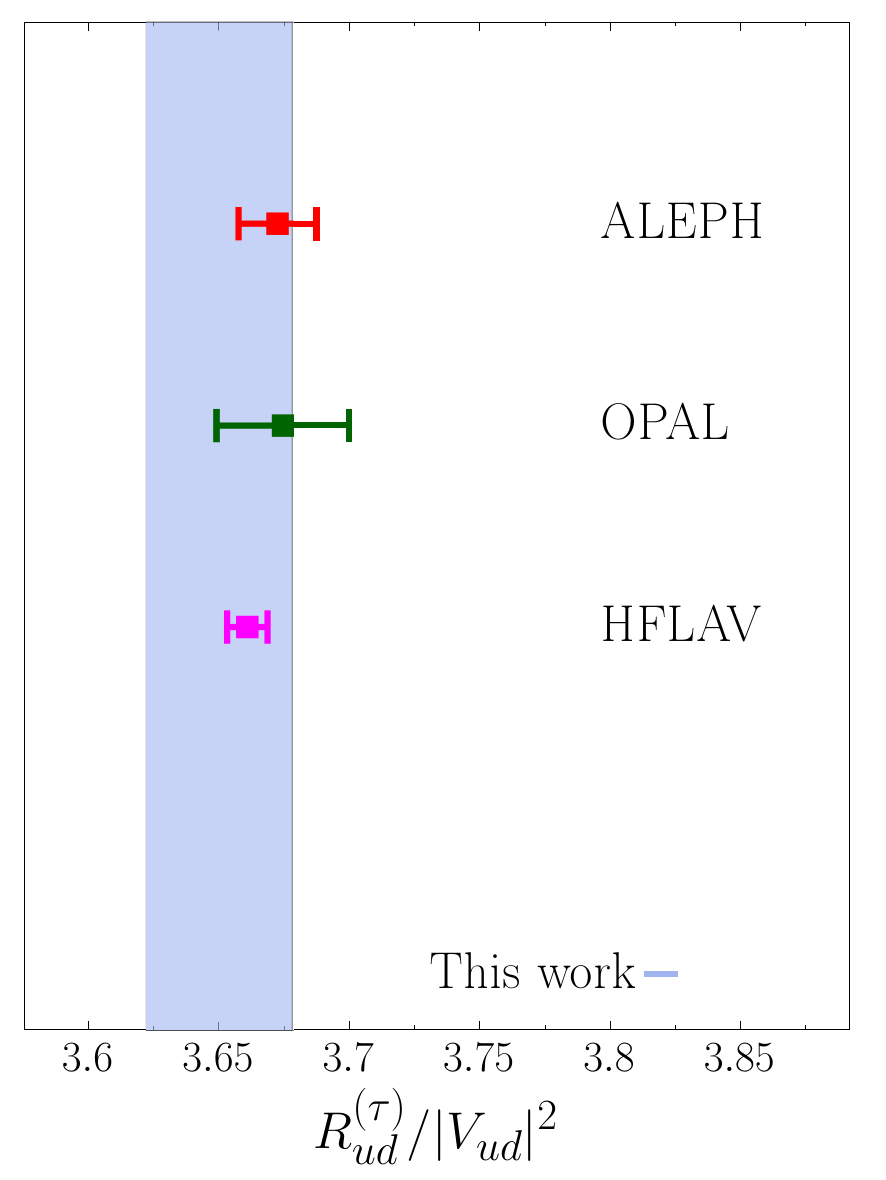}
 \vspace{-3mm}
 \caption{
   Left:
   differential decay rate $\bar{X}({\bf q}^2)$ as a function of ${\bf q}^2$
   (figure from Ref.~\cite{incl:Gambino+:22}).
   Symbols and bands in different colors show inclusive rate for two choices 
   of the weak current ($J_\mu = V_\mu$ or $A_\mu$) and its polarization
   (parallel or perpendicular to ${\bf q}$).
   These are compared with the ground state contribution
   from $B_s \to D_s^{(*)}\ell\nu$ shown by dashed lines.
   Right:
   comparison of $R_{ud}/|V_{ud}|^2$ (figure from Ref.~\cite{incl:tau:ETM}).
   The horizontal band represents ETM collaboration's lattice estimate.
   For experimental estimates (squares),
   $R_{ud}$'s from 
   ALEPH~\cite{tauBR:ALEPH}, OPAL~\cite{tauBR:OPAL}
   and HFLAV review~\cite{HFLAV21}
   are combined with $|V_{ud}|$ from the nuclear $\beta$ decays~\cite{NuclDepCorr}.
 }
\label{fig:incl:Bs}
\end{figure}
An extensive feasibility study of this apporach has been presented
in Ref.~\cite{incl:Gambino+:22}. The CPU cost is largely reduced
by focusing on the $B_s\!\to\!X_{cs}\ell\nu$ decay withtout valence light quarks
at a single combination of unphysicall heavy pion and light bottom quark masses.
The left panel of Fig.~\ref{fig:incl:Bs} shows a comaprison between
a differential ${\bf q}^2$ distribition of the inclusive rate $\bar{X}({\bf q}^2)$
(symbols and bands)
and corresponding ground state contribution from $B_s \to D_s^{(*)}\ell\nu$ decay
(dashed lines).
Their good consistency demonstrates the validiy of the approach to evaluate
the energy integral of the hadronic tensor using lattice four-point functions.
The authors also note that the excited state contribution can be suppressed
due to the limited phase space at the unphysically small bottom quark mass.

Toward practical application to determine CKM elements,
systematics of the approach are under intense investigation
in more realistic setups.
To evatulate the energy integral~(\ref{eqn:incl}) using $C_{\mu\nu}$ on the lattice,
the integration kernel $K_{\mu\nu}$ has to be approximated to a smooth function
of $\omega$~\cite{smearedSF,energyInt:GH,smearedSF:Hansen+,smearedSF:Bailas+}.
Reference~\cite{BGvsChev} studied systematics due to this approximation
using an HQET-based bottom quark action at its physical mass.
A study of FVEs, which are one of the most non-trivial systematic
uncertainties, was reported at this workshop~\cite{Kellermann:CKM23,Kellermann:Lat23}.
The authors focus on the $D_s \to X_{ss}\ell\nu$ decay
to control discretization errors and chiral extrapolation in valence quark masses.
In this safe setup,
they develop a model to describe FVEs by taking the contribution
for the two-body $K\bar{K}$ final states as an example.
Their model shows good consistency with their lattice data at ${\bf q}^2 = 0$.
Further studies at other values of ${\bf q}^2$
and direct simulations in different volumes
are necessary towards good control of FVEs.

Another interesing example of the inclusive anaysis is
ETM collaboration's study of the $\tau \to X_{ud}\nu$ decay
to determine $|V_{ud}|$~\cite{incl:tau:ETM}.
The corresponding hadornic tensor appears
in the spectral representation of the lattice two-point functions,
which are computationally much less expensive than the four-point functions
for the inclusive $B_{(s)}$ decays.
Therefore, both statistical and systematic uncertainties are under better contol:
namely, high statistics simulations with all relevant quark masses set to
their physical value in two different volumes to study FVEs
and at three lattice spacings to take the limit of $a = 0$.
Their estimate of the normalized inclusive decay rate
$R_{ud} = \Gamma(\tau \to X_{ud}\nu)/\Gamma(\tau \to e\bar{\nu}\nu)$
divided by $|V_{ud}|^2$ is in good agreement with experiment
as shown in the right panel of Fig.~\ref{fig:incl:Bs}.
They achieved 0.4\,\% determination of $|V_{ud}| = 0.9752(37)_{\rm th}(10)_{\rm ex}$,
which is nicely consistent with the precise (0.03\,\%) estimate 0.97373(31)
from the nuclear $\beta$ decays\cite{NuclDepCorr}.
The uncertainty is dominated by theoretical ones
from isopin corrections, statistics and FVEs.
With better control of these errors,
the inclusive $\tau$ decay would provide deteremination of $|V_{ud}|$
competitive to those from the neutron (0.1--0.2\,\%)
and pion (0.3\,\%) $\beta$ decays.
An interesting future direction is to extend the ETM study to $\tau \to X_{us}\nu$
to provide an alternative determination of $|V_{us}|$,
because a tension with the determinations from excluive kaon decays has been reported~\cite{HFLAV21}\footnote{
After the workshop,
the ETM collaboration performed
the determination of $|V_{us}|$~\cite{incl:tau:ETM:Vus},
which is consistent with the conventional determination~\cite{incl:tau:Gamiz+},
and hence confirms the tension with the kaon decays.}.


\section{Conclusions}

In this article,
we review status and progress of lattice QCD for the determination of CKM elements.
For gold-plated processes with light, strange and charm quarks,
independent realistic similations are becoming available leading to, for instance,
sub-\% determination of relevant hadronic inputs for the kaon (semi)leptonic decays.
More independent studies to crosscheck such precision calculations are highly welcome.
Strong isospin and radiative corrections are also under active investiations.

While simulations of bottom quarks are still limited to unphysically small masses
or effective-theory-based actions, many interesting quantites are being studied
including the $B \to D^*\ell\nu$ and $B_s$ decay form factors.
Systematics due to the unphysical setup are expected to be removed in the next decade.
We also note that close interplay between theorists and experimentalists is important
to resolve the long standing problem including $|V_{cb}|$ tension
and to maximaize the sensitivy of on-going experiments to new physics.

It is encouraging that wider applications are becoming avialable with
newly developed finite volume frameworks for
inclusive decays as well as
decays involving unstable particles, such as $B \to \rho\ell\nu$.
A practical application to determine $|V_{ud}|$ (and $|V_{us}|$)
from the inclusive $\tau$ decay has been already avilable.

\vspace{5mm}
\noindent
The work of TK is supported in part by JSPS KAKENHI Grant Numbers
JP21H01085 and JP22K21347, 
and by ``Program for Promoting Researches on the Supercomputer Fugaku''
(Simulation for basic science: approaching the new quantum era).


\bibliographystyle{amsplain}

\end{document}